\documentclass[letterpaper]{article} 
\usepackage{aaai2026}  

\usepackage{etoolbox}

\makeatletter
\def\copyright@text{} 
\makeatother

\usepackage{amsmath}
\usepackage{amsfonts}

\usepackage{times}  
\usepackage{helvet}  
\usepackage{courier}  
\usepackage[hyphens]{url}  
\usepackage{graphicx} 
\usepackage{natbib}  
\usepackage{caption} 
\usepackage{algorithm}
\usepackage{algorithmic}

\usepackage{subfigure}
\usepackage{amsthm}
\newtheorem{theorem}{Theorem}

\usepackage{amssymb}
\usepackage{multirow}
\usepackage{multicol}
\usepackage{booktabs}
\usepackage{colortbl}
\usepackage{pifont}
\usepackage{graphicx}
\usepackage{dsfont}

\usepackage{colortbl} 
\usepackage{array}

\usepackage{xspace}

\usepackage{tikz}
\usepackage{arydshln}
\newcommand{\our}{\textsc{ScaleFL}\xspace}

\usepackage[table,xcdraw]{xcolor}

\usepackage{mathtools}
\usepackage{pgfplots}

\usetikzlibrary{positioning,matrix,calc,arrows.meta,decorations.pathreplacing,shapes.geometric}
\definecolor{mygreen}{RGB}{46,139,87}
\definecolor{myred}{RGB}{255,152,150}
\definecolor{myblue}{RGB}{30,144,255}
\definecolor{myyellow}{RGB}{219,219,141}
\definecolor{mybrown}{RGB}{197,157,148}
\usepackage{pgfplotstable}

\usepackage{newfloat}
\usepackage{listings}
\DeclareCaptionStyle{ruled}{labelfont=normalfont,labelsep=colon,strut=off} 
\floatstyle{ruled}
\newfloat{listing}{tb}{lst}{}
\floatname{listing}{Listing}
\title{Bridging Memory Gaps: Scaling Federated Learning for Heterogeneous Clients}

\author {
    Yebo Wu\textsuperscript{\rm }, 
    Jingguang Li\textsuperscript{\rm },
    Chunlin Tian\textsuperscript{\rm },
    KaHou Tam\textsuperscript{\rm },
    Li Li\textsuperscript{\rm }\thanks{Corresponding author.},
    Chengzhong Xu\textsuperscript{\rm }
}
\affiliations {
    \textsuperscript{\rm }State Key Laboratory of Internet of Things for Smart City, University of Macau\\
    \{yc37926, mc45005, yc27402, yc37436, llili, czxu\}@um.edu.mo
}

\usepackage{bibentry}

\begin{document}

\maketitle

\begin{abstract}

Federated Learning (FL) enables multiple clients to collaboratively train a shared model while preserving data privacy. However, the high memory demand during model training severely limits the deployment of FL on resource-constrained clients. To this end, we propose \our, a scalable and inclusive FL framework designed to overcome memory limitations through sequential block-wise training. The core idea of \our is to partition the global model into blocks and train them sequentially, thereby reducing training memory requirements. To mitigate information loss during block-wise training, \our introduces a Curriculum Mentor that crafts curriculum-aware training objectives for each block to steer their learning process. Moreover, \our incorporates a Training Harmonizer that designs a parameter co-adaptation training scheme to coordinate block updates, effectively breaking inter-block information isolation. Extensive experiments on both simulation and hardware testbeds demonstrate that \our significantly improves model performance by up to 84.2\%, reduces peak memory usage by up to 50.4\%, and accelerates training by up to 1.9$\times$.

\end{abstract}
\section{Introduction}
\label{sec:intro}

Federated Learning (FL)~\cite{mcmahan2017communication} is a distributed learning paradigm that enables multiple clients to collaboratively train a shared model while preserving data privacy. 
Instead of gathering the raw data and conducting centralized training on the cloud, FL directly performs local training on participating clients and then aggregates the local updates~\cite{wang2025indoor,wang2023fedins2}. 
Despite the promising benefits, one fundamental and prevailing challenge that hinders the deployment of FL in real-world cases is the memory wall~\cite{wu2025memory,zhan2024heterogeneity}.
During local training, all intermediate activations, model weights, and optimizer states must be retained in memory~\cite{wu2025survey}, resulting in high memory footprint.
For instance, training a ResNet34 on ImageNet consumes more than 12 GB of memory. However, the available RAM on off-the-shelf mobile devices typically ranges from 4 to 12 GB~\cite{tian2024breaking}.
This means that these edge devices are unable to afford local training, thereby preventing their valuable data from contributing to the global model.
This situation is further exacerbated as model architectures are becoming deeper and wider to retrieve higher analysis capability~\cite{wu2025learning}.

Recently, several approaches have been proposed to resolve resource limitations in FL, which can be mainly divided into the following two categories: 1) model-heterogeneous training ~\cite{cho2022heterogeneous, zhang2022fedzkt} and 2) partial training~\cite{diao2020heterofl, yang2022partial}. 
Model-heterogeneous training customizes local models for each client based on their memory budgets and aggregates different local models through knowledge distillation~\cite{hinton2015distilling}.
However, the distillation process typically requires a high-quality public dataset, which is difficult to obtain in FL due to privacy concerns.
Partial training tailors the global model through width or depth scaling based on the resource constraints of participating clients, and then assigns the corresponding sub-models to them.
Specifically, width scaling~\cite{alam2022fedrolex} yields sub-models of varying complexity by adjusting the number of channels in convolutional layers. However, this strategy compromises the model architecture, leading to performance degradation. Similarly, depth scaling~\cite{kim2022depthfl, liu2022no} customizes sub-models with varying depths to handle memory constraints. Meanwhile, this mechanism restricts the global model’s trainable complexity, as the complexity is determined by the client with the largest memory capacity.

In response to these challenges, we propose \our, a scalable and inclusive FL framework that addresses the memory bottleneck through sequential block-wise training. Rather than training the entire model in each round, \our partitions the global model into smaller blocks and updates them sequentially, thereby minimizing memory requirements during local training. Each block contains a specified number of layers and corresponds to a distinct training stage. 
The process starts by training the first block, safely freezing it after convergence, and then triggering the training of the next one.
This process iterates until all blocks have been fully trained. In this way, the memory space required for storing intermediate activations and gradients of the frozen blocks can be effectively released. Thus, more clients can meet the memory requirements and contribute to the global model with their private data. 
However, this sequential training paradigm introduces two critical challenges, which motivate the design of our core techniques.

\textbullet\textbf{\textit{Challenge 1: Information Loss in Each Stage.}}
Training each block solely based on the task-specific objective drives it to extract high-level features that benefit short-term goals, while neglecting its role in the feature extraction process of the entire model.
With limited parameter capacity, each block struggles to retain valuable information required by downstream blocks, resulting in significant information loss. These degraded representations hinder subsequent blocks from extracting more critical features, ultimately impairing overall model performance.
To address this challenge, we introduce the Curriculum Mentor, which crafts curriculum-aware training objectives for each block,  grounded in information bottleneck theory, to steer their optimization process. By encouraging each block to learn structured feature representations, the Curriculum Mentor effectively mitigates information loss during block-wise training.

\textbullet\textbf{\textit{Challenge 2: Information Isolation across Blocks.}} Another critical challenge is the information isolation across blocks. Information isolation arises for the following two reasons: 1) Later blocks initiate training only after the preceding blocks have converged and been frozen, thereby preventing them from adapting to the evolving features of earlier blocks. 2) Since each block is updated independently, the absence of gradient flow across blocks leads to fragmented learning and gradient misalignment.
To address this challenge, we propose the Training Harmonizer, which implements a parameter co-adaptation training scheme to facilitate bidirectional information flow across blocks.
Specifically, it adaptively grows the model in each round, allowing later blocks to align with the update trajectories of preceding blocks and make timely adjustments, thereby enhancing forward information flow.
Moreover, to facilitate inter-block gradient flow during backpropagation, it concurrently trains each block alongside the final layers of its preceding block.

We evaluate the effectiveness of \our on multiple representative datasets. Compared to state-of-the-art methods, \our achieves up to 84.2\% improvement in model accuracy, reduces peak memory usage by up to 50.4\%, and delivers up to 1.9$\times$ speedup in convergence.
The main contributions of this work are summarized as follows:

\begin{itemize}
    \item We propose \our, a scalable and inclusive FL framework that significantly reduces training memory requirements through sequential block-wise training, enabling resource-limited devices to participate effectively.

    \item We introduce two core components,
    the Curriculum Mentor and the Training Harmonizer, to support this training paradigm. They synergistically shape the learning behavior of each block, facilitating coherent and structured feature learning.
    
    \item We conduct extensive experiments on multiple benchmark datasets to demonstrate the effectiveness and robustness of \our.

\end{itemize}

\section{Background and Motivation}

\subsection{The Sequential Block-Wise Training Paradigm}\label{sec_paradigm}

\begin{figure}[t]
  \centering
  \includegraphics[width=0.9\linewidth]{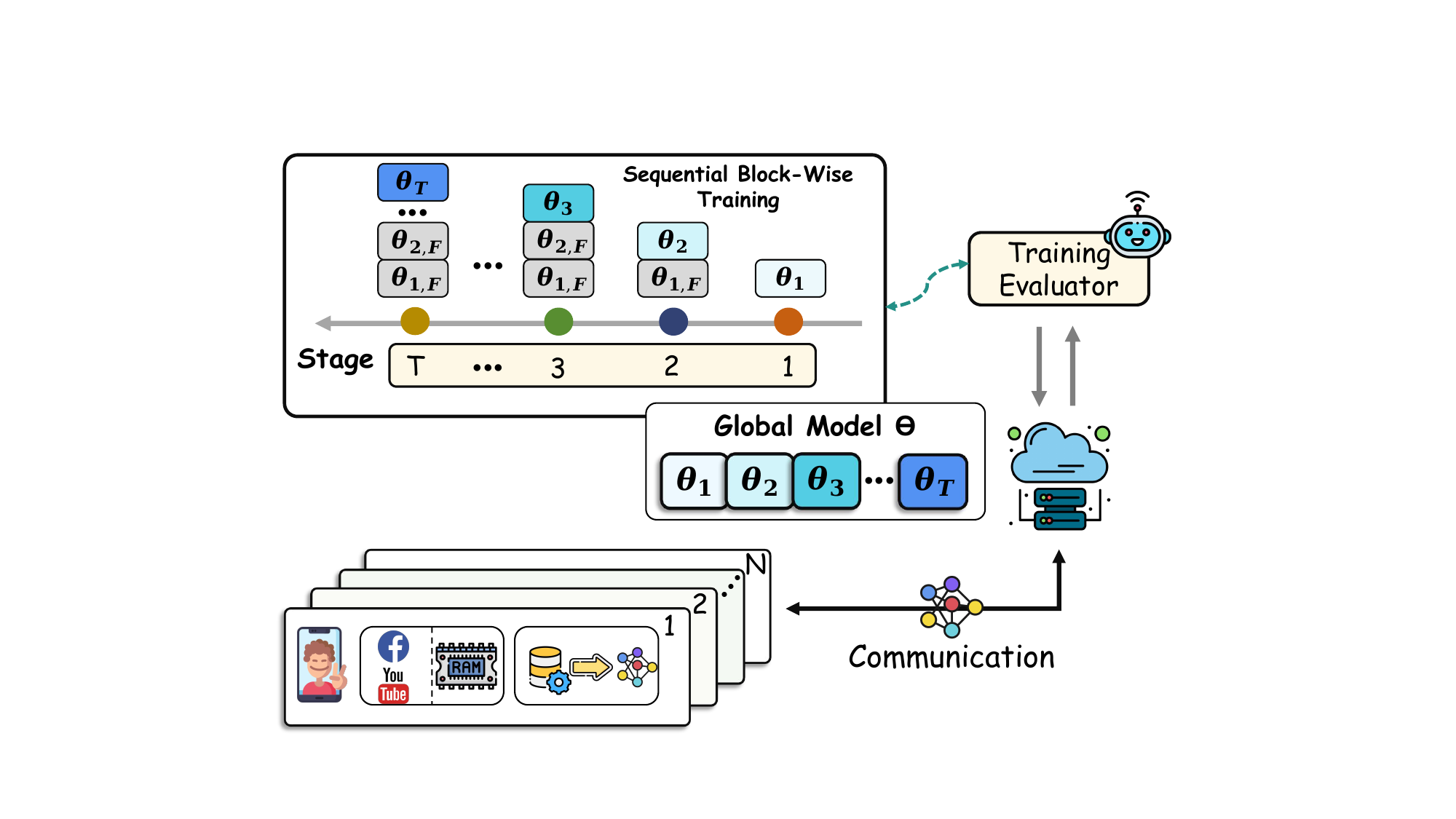}
  \caption{Workflow of the sequential block-wise training.}
  \label{motivation2}
  \vspace{-10pt}
\end{figure}

In an FL system comprising \(N\) clients, each client \(n\) possesses its own local dataset, denoted as \(D_n\), where \(n \in [1, N]\).
Instead of updating the entire model in each round, the global model $\Theta$ is initially partitioned into $T$ disjoint blocks, denoted as $\Theta = [\theta_{1}, \theta_{2}, \ldots, \theta_{T}]$.
Each block corresponds to a distinct training stage, and a block that has been frozen is denoted as $\theta_{t,F}$.
Since all blocks except the last one lack output modules, they cannot complete the training procedure independently. To resolve this, we concatenate an output module $\theta_{Op}$ to each block to ensure independent training.
Figure~\ref{motivation2} illustrates the workflow of the sequential block-wise training paradigm. 
The overall training process consists of the following steps~\cite{wu2024heterogeneity}: 
\begin{itemize}
    \item 1) \textbf{Model Construction:} The server constructs a sub-model for the current training stage $t$ (where $t \in [1, T]$), defined as $\Theta_{g,t} = [\theta_{1,F}, \theta_{2,F}, \ldots, \theta_t, \theta_{\text{Op}}]$.

    \item 2) \textbf{Local Training:} The constructed sub-model is broadcast to the selected client set \textit{S}, where local training is performed. Since only block $\theta_t$ and the output module $\theta_{\text{Op}}$ are updated, these parameters are uploaded to the central server for aggregation.

    \item 3) \textbf{Model Aggregation:} The central server aggregates the received updates using Equation~\eqref{aggre}, where $|D|$ denotes the total data size of the participating clients.

    \begin{equation}
        \label{aggre}
        [\theta_{g,t}^{r},\theta_{g,Op}^{r}] = \sum_{n \in S} \frac{|D_{n}|}{|D|}([\theta_{n,t}^{r},\theta_{n,Op}^{r}]).
    \end{equation}
    
    \item 4) \textbf{Progress Evaluation:} The server monitors the training process of block $\theta_{t}$ and determines whether convergence has been achieved.

    \item 5) \textbf{Model Growing:} 
    The server freezes the converged block and concatenates a new block $\theta_{t+1}$, designating it as the new global model for stage $t+1$ ($\Theta_{g,t+1} =[\theta_{1,F},\theta_{2,F},...,\theta_{t,F},\theta_{t+1},\theta_{Op}]$). 
\end{itemize}

These steps iterate until all blocks are fully trained. By updating only one block per round, this training paradigm significantly reduces the peak memory footprint on participating clients.

\subsection{Limitations of Sequential Block-Wise Training}\label{sec_limitations}

We then conduct experiments to evaluate the feasibility of applying the sequential block-wise training paradigm in real-world scenarios. For benchmarking, we compare \our against two baseline methods: 1) FedAvg~\cite{mcmahan2017communication}, and 2) a theoretical baseline, denoted as \textsc{TheoFL}, which assumes that all participating clients have sufficient memory resources to perform local training. We establish a federated environment with 100 mobile devices, assigning available memory to each device based on real-world profiling traces~\cite{wu2024heterogeneity}. The CIFAR10 and CIFAR100 datasets are distributed across devices under both IID and non-IID settings, with ResNet18 used as the global model.

Figure~\ref{fig:limitation} shows that FedAvg suffers from significant performance degradation compared to \textsc{TheoFL}, with accuracy drops of 11.6\% on CIFAR10 and 22.5\% on CIFAR100 under the IID setting. 
This performance degradation arises because many edge devices lack sufficient memory to support local training. As a result, their valuable data cannot contribute to the global model. These results underscore that addressing the memory constraints is critical for the practical deployment of FL. 
Moreover, although the sequential block-wise training paradigm partially narrows the performance gap between FedAvg and \textsc{TheoFL}, it still exhibits a 6.7\% accuracy reduction on CIFAR10 and 5.2\% on CIFAR100 under the IID setting.
This suggests that naïvely applying the sequential block-wise training paradigm is insufficient, highlighting the need for more principled designs.

\definecolor{red}{RGB}{172,21,28}
\definecolor{blue}{RGB}{39,89,167}
\definecolor{red1}{RGB}{203,104,104}
\definecolor{blue1}{RGB}{104,155,203}
\definecolor{color1}{RGB}{235,164,122}
\definecolor{color3}{RGB}{78,172,183}
\definecolor{color2}{HTML}{95a792}

\begin{figure}[t]
    \centering
    \vspace{1.0mm}
\begin{tikzpicture}
    \scriptsize{

  \begin{axis}[
    at={(-5.5em,-15.5em)},
    anchor=south west,
    ymajorgrids,
    grid style=dashed,
    legend style={at={(0.04,1)}, anchor=south west},
    legend cell align={left},
    ybar,
    enlarge x limits=0.5,
    xtick align=inside,
    height=.23\textwidth,
    width=.27\textwidth,
    bar width=1.3em,
    xlabel={\scriptsize{(a) CIFAR10.}},
    xlabel style={scale=1.2, yshift=0.8em, xshift=0.1em},
    ylabel=\footnotesize{\scriptsize Accuracy (\%)},
    ylabel style={scale=1.2, yshift=0.3em},
    symbolic x coords={{1}, {2},},
    xtick=data,
    ymin=75,
    ymax=92,
    ytick={75,80,85,90,95},
    nodes near coords align={vertical},
    xticklabels={IID, non-IID},
    ylabel style={yshift=-2em},
    yticklabel style={/pgf/number format/fixed,/pgf/number format/fixed zerofill,/pgf/number format/precision=0,rotate=0,scale=1.0},
    legend style={yshift=0.2em,xshift=4.2em,font={\tiny},cells={anchor=west},fill opacity=0.8, scale=1.0, legend columns=3}
    ]
    \addplot[fill=color1, draw=color1, area legend] coordinates {({1},77.9) ({2},76.8)};
    \addlegendentry{\scalebox{1.0}{{FedAvg}}}
    \addplot[fill=color2, draw=color2, area legend] coordinates {({1},82.8) ({2},78.2)};
    \addlegendentry{\scalebox{1.0}{\textsc{SBT}}}
    \addplot[fill=color3, draw=color3, area legend] coordinates {({1},89.5) ({2},88.4)};
    \addlegendentry{\scalebox{1.0}{\textsc{TheoFL}}}
  \end{axis}
	
  \begin{axis}[
    at={(11.5em,-15.5em)},
    anchor=south west,
    ymajorgrids,
    grid style=dashed,
    legend style={at={(0.02,1)}, anchor=south west},
    legend cell align={left},
    ybar,
    enlarge x limits=0.5,
    xtick align=inside,
    height=.23\textwidth,
    width=.27\textwidth,
    bar width=1.3em,
    xlabel={\scriptsize{(b) CIFAR100.}},
    xlabel style={scale=1.2, yshift=0.8em, xshift=0.1em},
    ylabel=\footnotesize{\scriptsize Accuracy (\%)},
    ylabel style={scale=1.2, yshift=0.3em},
    symbolic x coords={{1}, {2},},
    xtick=data,
    ymin=30,
    ymax=63,
    ytick={30,40,50,60},
    nodes near coords align={vertical},
    xticklabels={IID, non-IID},
    ylabel style={yshift=-2em},
    yticklabel style={/pgf/number format/fixed,/pgf/number format/fixed zerofill,/pgf/number format/precision=0,rotate=0,scale=1.0},
    legend style={yshift=0.2em,xshift=4.2em,font={\tiny},cells={anchor=west},fill opacity=0.8, scale=1.0, legend columns=3}
    ]
    \addplot[fill=color1, draw=color1, area legend] coordinates {({1},37.1) ({2},35.2)};
    \addplot[fill=color2, draw=color2, area legend] coordinates {({1},54.4) ({2},48.5)};
    \addplot[fill=color3, draw=color3, area legend] coordinates {({1},59.6) ({2},58.1)};
  \end{axis}
}   
\end{tikzpicture}
\vspace{-5mm}
\caption{
Performance comparison of FedAvg, \textsc{TheoFL}, and the sequential block-wise training paradigm (denoted as \textsc{SBT}) on CIFAR10 and CIFAR100.
}
\label{fig:limitation}
\vspace{-4mm}
\end{figure}

This performance gap stems from two fundamental challenges.
\textit{\textbf{Challenge 1: Information Loss in Each Stage.}}
Training each block solely based on the task-specific objective leads to the loss of valuable information that are critical for subsequent blocks. \textit{How can we formulate block-specific training objectives to mitigate stage-wise information loss?} \textit{\textbf{Challenge 2: Information Isolation across Blocks.}}
Sequentially training each block to convergence hinders information propagation across blocks. 
\textit{How can we alleviate this isolation and promote effective inter-block collaboration?}

\newcommand{\E}{\mathbb{E}}
\newcommand{\norm}[1]{\left\lVert#1\right\rVert}

\section{\our}

Motivated by these findings, we propose \our, a framework that incorporates two core components—the Curriculum Mentor and the Training Harmonizer—to systematically guide the training process of each block and address the core challenges in sequential block-wise training.


\subsection{Curriculum Mentor}\label{section4_1}

\begin{figure*}[!t]
  \centering
  \includegraphics[width=0.95\linewidth]{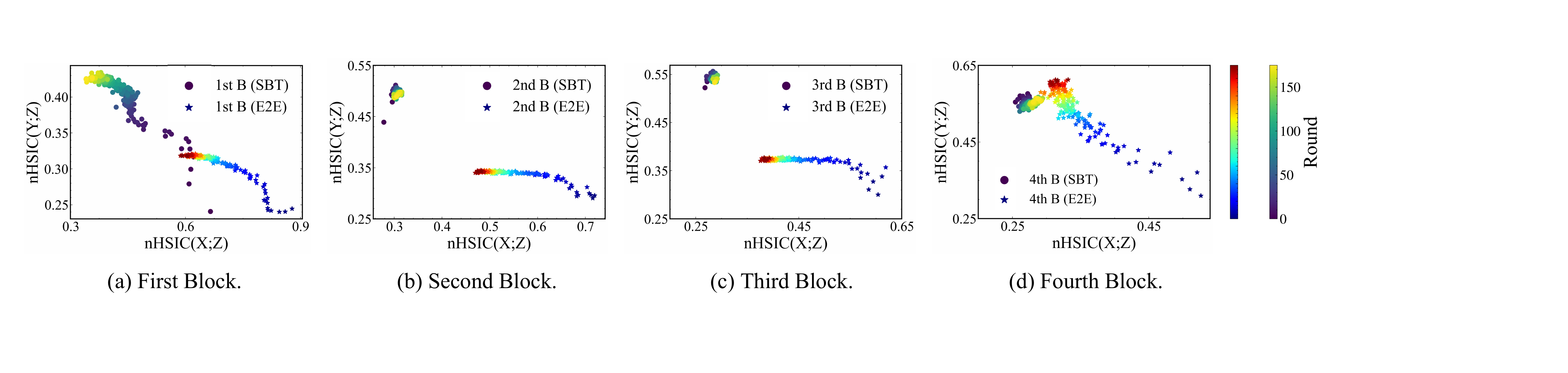}
  \vspace{-5pt}
  \caption{nHSIC plane dynamics for different blocks of ResNet18 trained on CIFAR10, with the model divided into four blocks. The color gradation shows the training progress, i.e., the number of rounds.}
  \label{Hsic_plane}
  \vspace{-12pt}
\end{figure*}

To address the challenge of information loss, \our introduces the Curriculum Mentor, which crafts curriculum-aware training objectives for each block. 
Analogous to well-designed human curricula~\cite{wang2021survey}, these objectives are organized from simple to complex, effectively guiding each block to learn structured feature representations and mitigating information loss at each stage.

However, modulating feature extraction at each stage is challenging due to the lack of fine-grained control over which features are retained or discarded by each block.
To address this, we leverage the information bottleneck (IB) theory~\cite{tishby2000information} for a deeper analysis. IB theory seeks to find an optimal intermediate representation $Z$ between a given input $X$ and output $Y$, preserving task-relevant information while eliminating task-irrelevant details. This objective is formalized as minimizing the following function:
\begin{equation}
    \label{IB}
    I(X; Z)-\beta I(Y; Z),
\end{equation}
where \( I(X; Z) \) and \( I(Y; Z) \) denote the mutual information between the input and representation, and between the output and representation, respectively, and \( \beta \) is a hyperparameter that balances information retention and the relevance of extracted features.
For an $L$-layer model, each layer collaboratively processes the data, generating a series of intermediate representations $\{Z_{1}, Z_{2},...Z_{L}\}$. 
This hierarchical transformation can be modeled as a Markov chain~\cite{yu2023go}, satisfying:
\begin{equation}
    \label{IB_ineqa}
    I(Y; X) \geq I(Y; Z_{1}) \geq \cdots \geq I(Y; Z_{L}).
\end{equation}

As shown in Equation~\eqref{IB_ineqa}, each layer’s processing leads to a gradual loss of task-relevant information. This effect is even more pronounced in the sequential block-wise training paradigm, which we will discuss further below. To counteract this, we can integrate mutual information into the training objective to mitigate task-relevant information loss at each stage.
However, due to the high dimensionality of the activations, directly computing mutual information results in significant computational overhead~\cite{sakamoto2024end}.
To address this, we employ the Hilbert-Schmidt Independence Criterion (HSIC) bottleneck~\cite{ma2020hsic, wang2021learning}, a statistical measure that quantifies the dependence between variables, to estimate mutual information. Since HSIC only requires computing sample similarity matrices via kernel functions, it significantly reduces computational complexity. The detailed formulation of HSIC is provided in the appendix.
Specifically, we use \( \text{nHSIC}(X; Z) \) and \( \text{nHSIC}(Y; Z) \) to estimate \( I(X; Z) \) and \( I(Y; Z) \), respectively, where nHSIC is defined by the Hilbert-Schmidt norm of the normalized cross-covariance operator~\cite{gretton2005measuring}:
\begin{equation}
\small
\begin{split}
    \text{nHSIC}(X; Z) &= \frac{\text{HSIC}(X; Z)}{\sqrt{\text{HSIC}(X; X) \cdot \text{HSIC}(Z; Z)}}, \\
    \text{nHSIC}(Y; Z) &= \frac{\text{HSIC}(Y; Z)}{\sqrt{\text{HSIC}(Y; Y) \cdot \text{HSIC}(Z; Z)}}.
\end{split}
\label{eq_hsic}
\end{equation}

In this context, a higher \( \text{nHSIC}(Y; Z) \) suggests that \( Z \) captures more task-relevant features, while a higher \( \text{nHSIC}(X; Z) \) indicates that \( Z \) preserves more input-specific details.
To illustrate the information loss associated with block-wise training, we conduct experiments using ResNet18 on  CIFAR10, with the model divided into four blocks. For each block, we measure \( \text{nHSIC}(X; Z) \) and \( \text{nHSIC}(Y; Z) \) under two training approaches: end-to-end training (E2E) and sequential block-wise training (SBT).

Figure~\ref{Hsic_plane} presents the nHSIC plane dynamics for different blocks, with the X-axis representing \( \text{nHSIC}(X; Z) \) (input retention) and the Y-axis representing \( \text{nHSIC}(Y; Z) \) (task relevance). Figure~\ref{Hsic_plane}(a) shows that after  training the first block, the E2E approach yields a lower $\text{nHSIC}(Y; Z)$ but maintains a higher $\text{nHSIC}(X; Z)$, indicating that it preserves substantial input information. In contrast, SBT achieves a higher $\text{nHSIC}(Y; Z)$ but at the cost of significantly reducing $\text{nHSIC}(X; Z)$, as it discards considerable input information in favor of task-specific features. Moving to Figure~\ref{Hsic_plane}(b), \ref{Hsic_plane}(c), and \ref{Hsic_plane}(d), which show the subsequent blocks' training process, we observe that SBT’s $\text{nHSIC}(Y; Z)$ remains stagnant. This stagnation arises from the loss of valuable input information during training the first block, which hinders later blocks from extracting more critical features. Conversely, in E2E training, $\text{nHSIC}(Y; Z)$ gradually increases across blocks, eventually surpassing SBT and achieving superior performance.
This analysis reveals that training each block solely based on the task-specific objective leads to substantial information loss, ultimately compromising model performance.

Furthermore, according to the principle of the inverse data processing inequality~\cite{yu2023go}, \( I(Y; Z) \leq I(X; Z) \), meaning that to increase \( I(Y; Z) \), we must increase the lower bound \( I(X; Z) \). Building on these insights, we formulate the following curriculum-aware training objective for block \( t \):
\begin{equation}
    \small
    \label{IB_loss_old}
    \mathcal{L}_{\mathbf{\theta}_{t}} = \mathcal{L}_{CE} - \lambda_{t} \cdot \text{nHSIC}(X; Z_{t}) - \gamma_{t} \cdot \text{nHSIC}(Y; Z_{t}),
\end{equation}
where \( \mathcal{L}_{CE} \) denotes the cross-entropy loss, and \( \lambda_{t} \) and \( \gamma_{t} \) are hyperparameters that balance the contributions of each term.
When setting values for \( \lambda_{t} \) and \( \gamma_{t} \), we start with a higher \( \lambda_{t} \) and a lower \( \gamma_{t} \) in the initial blocks to emphasize input information retention. As training progresses, \( \lambda_{t} \) is gradually decreased while \( \gamma_{t} \) is gradually increased, allowing the model to shift its focus toward task-specific feature extraction in later blocks.

It is worth noting that \our maintains compatibility with existing FL approaches, as each training stage functions in a similar manner to conventional FL. 
To address data heterogeneity in FL, we incorporate L2 regularization into Equation~\eqref{IB_loss_old}, yielding the following curriculum-aware training objective at round $r$:
\begin{equation}
    \small
    \label{IB_loss2}
    \mathcal{L}_{t}^{r} = \mathcal{L}_{\mathbf{\theta}_{t}} +\frac{\mu}{2}|| \theta_{t}^{r} - \theta_{t}^{r-1} ||_{2}^{2},
\end{equation}
where $\mu$ is a hyperparameter and $\theta_{t}^{r-1}$ represents the parameters of block $t$ from the round $r-1$. In contrast to previous work~\cite{li2020federated, zhang2022fedzkt}, 
we apply regularization only to the block currently being trained, rather than to all model parameters.

\subsection{Training Harmonizer}

\begin{figure}[!t]
  \centering
  \includegraphics[width=0.9\linewidth]{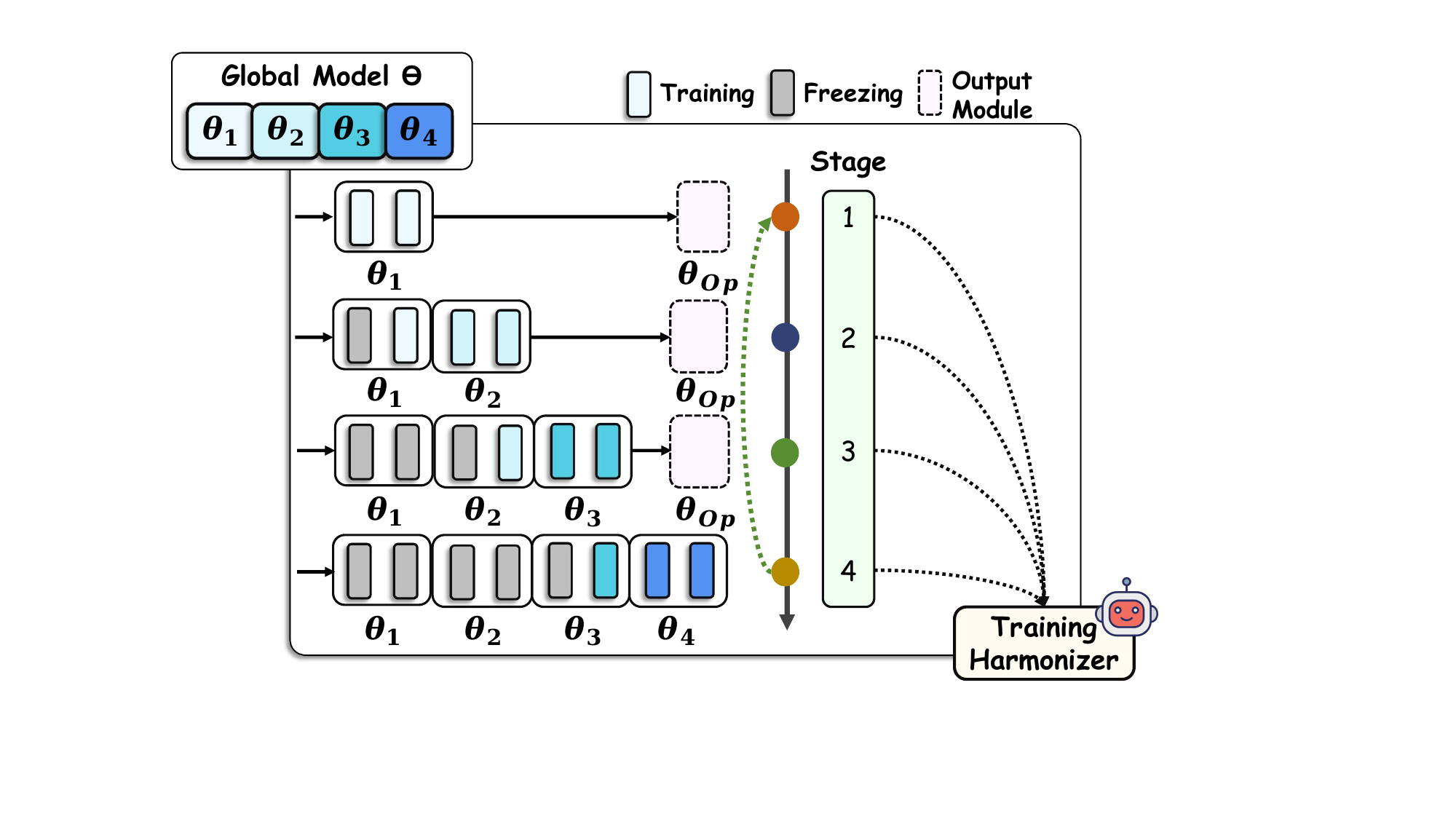}
  \caption{Illustration of the Training Harmonizer, which coordinates block updates through a parameter co-adaptation scheme. The global model is partitioned into four blocks, each composed of two layers.}
  \label{design2}
  \vspace{-15pt}
\end{figure}

To address the challenge of information isolation, \our introduces the Training Harmonizer, which implements a parameter co-adaptation training scheme to enhance information interaction across blocks during both forward and backward propagation.

Information isolation primarily stems from two factors:
1) \textbf{Restricted forward information flow:} During sequential block-wise training, downstream blocks commence training only after earlier blocks have converged. 
This delay prevents later blocks from promptly adapting to the evolving representations learned by preceding blocks, disrupting forward information flow. 2) \textbf{Restricted backward information flow:} Since gradients are confined within each block, backward information flow is restricted, resulting in gradient isolation. This isolation impairs the model’s ability to capture inter-block dependencies, leading to fragmented learning and gradient misalignment.

The parameter co-adaptation training scheme effectively fosters information interaction during both forward and backward propagation. Specifically, the Training Harmonizer dynamically orchestrates the learning process of each block in each round to enhance forward information flow. For example, as shown in Figure~\ref{design2}, the first block $\theta_{1}$ is trained in the first round, and then the second block $\theta_{2}$ is introduced and trained in the second round.
Once all blocks have been incorporated, the Training Harmonizer re-trains the first block, enabling downstream blocks to continuously align with the evolving updates of preceding blocks in real time. 
Additionally, the Training Harmonizer employs a concurrent training strategy to facilitate gradient flow during backward propagation.
Specifically, the later layers of the preceding block are co-trained with the current block, allowing gradients to propagate more freely between blocks. 
For instance, when training the second block, the later layers of the first block are simultaneously updated, as illustrated in Figure~\ref{design2}.
We define the layers designed to break gradient isolation as connecting layers, denoted as $L_{t}$ for block $t$. The updating process for block $t$ at the $k$-th step can be formulated as:
\begin{equation}
    \small
    \label{break_gradient}
    \theta_{n,t}^{k+1}+L_{n,t-1}^{k+1} \leftarrow (\theta_{n,t}^{k}+L_{n,t-1}^{k})-\eta \cdot \frac{\partial \mathcal{L}_{n,t}^{k}}{\partial(\theta_{n,t}^{k}+L_{n,t-1}^{k})},
\end{equation}
where $\eta$ denotes the learning rate.
Algorithm~\ref{algorithm_neu} outlines the workflow of \our, where \textit{Line 4} enhances forward information flow and \textit{Line 8} reduces information loss while breaking gradient isolation. Moreover, we provide the convergence analysis of \our in the appendix.

\begin{algorithm}[!t]
\caption{The workflow of \our}
\label{algorithm_neu}
\begin{algorithmic}[1]
\REQUIRE Global model $\Theta$, $R$ rounds
\ENSURE Optimized global model $\Theta^{*}$
\STATE Divide $\Theta$ into $T$ blocks ($[\theta_{1}, \theta_{2}, \dots, \theta_{T}]$)

\STATE Construct output module $\theta_{Op}$ for each block

\FOR{$r = 1$ to $R$}

    \STATE Stage $t = r \mod T$

    \STATE Construct model $\Theta_{g,t} = [\theta_{1,F}, \theta_{2,F}, \dots, \theta_{t}, \theta_{Op}]$
    \STATE Select a set of clients $S$ based on memory capacity
    
    \FOR{each client $n$ in $S$}
        \STATE Perform local training using Equations~\eqref{IB_loss2} and~\eqref{break_gradient}

        \STATE Upload parameters $[L_{n,t-1}^{r}, \theta_{n,t}^{r}, \theta_{n,Op}^{r}]$
    \ENDFOR
    
    \STATE Server performs parameter aggregation:
    \small
    \begin{equation*}
        \label{eq_aggre}
        [L_{g,t-1}^{r}, \theta_{g,t}^{r}, \theta_{g,Op}^{r}] = \sum_{n \in S} \frac{|D_{n}|}{|D|}([L_{n,t-1}^{r}, \theta_{n,t}^{r}, \theta_{n,Op}^{r}]).
    \end{equation*}
\ENDFOR
\end{algorithmic}
\end{algorithm}
\vspace{-5pt}

\section{Evaluation}

\subsection{Experimental Setup}\label{sub_exp_set}

\begin{table*}[!h]
  \small
  \centering
  \setlength{\tabcolsep}{2.9pt} 
  \begin{tabular}{lccccccccccccc}
    \bottomrule[1.5pt]
    \addlinespace[2pt] 

    \multirow{2}{*}{\textbf{Method}} & \multirow{2}{*}{\textbf{Inclusive?}} & \multicolumn{3}{c}{\textbf{CIFAR10 (Non-IID)}} & \multicolumn{3}{c}{\textbf{CINIC10 (Non-IID)}} & \multicolumn{3}{c}{\textbf{CIFAR100 (Non-IID)}} & \multicolumn{3}{c}{\textbf{PR}} \\
    \cmidrule(lr){3-5} \cmidrule(lr){6-8} \cmidrule(lr){9-11} \cmidrule(lr){12-14}
    & & \textbf{Res18.} & \textbf{VGG11.} & \textbf{SqNet.} & \textbf{Res18.} & \textbf{VGG11.} & \textbf{SqNet.} & \textbf{Res18.} & \textbf{VGG11.} & \textbf{SqNet.} & \textbf{Res18.} & \textbf{VGG11.} & \textbf{SqNet.} \\
    \toprule[0.75pt]

    AllSmall &Yes&69.5\%&75.1\%&49.6\%  &67.8\%&69.2\%&32.1\%  &17.4\%&42.6\%&17.6\%&100\%&100\%&100\%  \\
    ExclusiveFL&No&76.8\%&79.3\%&40.6\% &69.0\%&67.1\%&51.8\% &35.2\%&46.9\%&22.4\%&18\%&22\%&11\%  \\

    DepthFL&No&65.1\%&75.0\%&49.7\% &61.2\%&63.2\%&43.2\% &33.3\%&40.4\%&23.5\%&43\%&43\%&39\%   \\
    HeteroFL&Yes&76.7\%&76.6\%&35.7\% &66.5\%&68.5\%&40.7\% &34.8\%&45.0\%&10.8\%&100\%&100\%&100\%   \\
    
    FedRolex&Yes&76.6\%&74.9\%&16.8\% &64.2\%&67.0\%&15.9\% &35.7\%&45.3\%&2.1\%&100\%&100\%&100\%   \\

    SmartFreeze & Yes & 76.7\% & 83.2\% & 52.5\% & 69.5\% & 69.8\% & 53.6\% & 48.2\% & 52.3\% & 26.7\% & 100\% & 100\% & 100\% \\
    ProFL & Yes & 78.2\% & 85.1\% & 55.3\% & 71.2\% & 70.8\% & 55.1\% & 49.6\% & 56.7\% & 28.3\% & 100\% & 100\% & 100\% \\

    \midrule
    \our & Yes&\textbf{80.4\%}&\textbf{87.6\%}&\textbf{58.0\%}&\textbf{74.1\%}&\textbf{72.3\%}&\textbf{57.0\%} &\textbf{51.2\%}&\textbf{62.0\%}&\textbf{31.2\%}&100\%&100\%&100\% \\
    \bottomrule[1.5pt]
  \end{tabular}
  \vspace{-4pt}
  \caption{Overall performance comparison of \our against baseline methods. $PR$ denotes the device participation rate, and $SqNet$ refers to SqueezeNet. \textbf{Bold} values indicate the best results.}
  \label{performance_total} 
  \vspace{-13pt}
\end{table*}

\noindent\textbf{Models and Datasets.} To comprehensively evaluate the effectiveness of \our, we adopt a suite of models commonly deployed on mobile devices~\cite{tam2024towards}, including ResNet18, ResNet34, VGG11\_BN, SqueezeNet~\cite{iandola2016squeezenet}, and the Vision Transformer (ViT)~\cite{dosovitskiy2020image}. 
Our experiments cover both IID and non-IID settings across five benchmark datasets: CIFAR10, CINIC10, CIFAR100, Mini-ImageNet~\cite{vinyals2016matching}, and FEMNIST~\cite{caldas2018leaf}, a large-scale FL dataset. The non-IID data partitioning follows a Dirichlet distribution with a concentration parameter $\alpha = 1$~\cite{tam2023federated}.

\noindent\textbf{Default Settings.} 
Our experiments are conducted using a hybrid setup of simulation and real-device testbeds~\cite{tian2022harmony,tam2024fedhybrid}. 
We deploy an on-device FL system with 100 heterogeneous mobile devices, and in each round, 10\% of them are randomly selected to participate.
During local training, SGD is employed as the optimizer with a weight decay of 5e-4, and the local epoch is set to 5. 
Each device is assigned a memory budget randomly sampled from the range 100–1000 MB, based on hardware profiling results~\cite{wu2024heterogeneity}, to ensure fair comparisons.
The hyperparameter $\mu$ is set to either 0.01 or 0.001, while the regularization coefficients $\lambda$ and $\gamma$ are tuned within the range $[0, 10]$. Additionally, the number of connecting layers, used to break gradient isolation, is configured to 2.

\noindent\textbf{Baselines.}
The following baselines are employed for evaluation purposes: 1) AllSmall~\cite{wu2025breaking} serves as a naïve baseline that downsizes the global model based on the memory of the weakest device, allowing all devices to participate.
2) ExclusiveFL~\cite{liu2022no} restricts participation to devices with sufficient memory to support local training.
3) DepthFL~\cite{kim2022depthfl} applies depth scaling to the global model to accommodate heterogeneous devices. 
4) HeteroFL~\cite{diao2020heterofl} employs static channel scaling in convolutional layers to address memory constraints.
5) FedRolex~\cite{alam2022fedrolex} dynamically performs width scaling on the global model through a sliding window. 
6) SmartFreeze~\cite{wu2024heterogeneity} simply employs the sequential block-wise training paradigm to reduce memory usage during local training.
7) ProFL~\cite{wu2025breaking} decomposes the sequential training process into shrinking and growing stages to improve block learning.

\subsection{Main Results}

Table~\ref{performance_total} presents the experimental results under the non-IID scenario, showing that \our consistently outperforms all baselines across various experimental settings.

Specifically, on CIFAR10, \our achieves accuracy improvements of 10.9\%, 12.5\%, and 8.4\% over AllSmall across the three models. This performance gap stems from the fact that AllSmall determines the global model based on the device with the lowest memory capacity, which significantly limits its feature extraction capability.
ExclusiveFL restricts participation to high-memory devices, leading to low device participation rates of 18\%, 22\%, and 11\%. In contrast, \our effectively leverages data from low-memory devices, resulting in accuracy improvements of 3.6\%, 8.3\%, and 17.4\%, respectively. 
\our achieves 15.3\%, 12.6\%, and 8.3\% higher accuracy than DepthFL across the three models, largely due to DepthFL’s insufficient engagement of low-memory devices, evidenced by participation rates of just 43\%, 43\%, and 39\%.
Although HeteroFL and FedRolex attain full device participation through width scaling, they compromise model architecture, especially for compact networks like SqueezeNet with limited channels. Consequently, \our achieves superior accuracy, with gains ranging from 3.7\% to 41.2\%.
Moreover, sequential block-wise training methods such as SmartFreeze and ProFL fail to address critical issues, including information loss and inter-block information isolation, resulting in accuracy drops of up to 5.5\% relative to \our.

Additionally, \our consistently outperforms all baselines on the CINIC10 dataset, achieving accuracy improvements ranging from 1.5\% to 41.1\%.
It also performs well on the more challenging CIFAR100 dataset, with gains ranging from 1.6\% to 33.8\%. These significant improvements stem from two key factors: 1) \our is an inclusive framework that leverages data from all devices, and 2) it incorporates two core components that effectively guide the block-wise training process.

\subsection{Generalization Analysis}

\begin{table}[!t]
  \small
  \centering
  \setlength{\tabcolsep}{2.6pt} 

  \begin{tabular}{lccccccccccc}

    \bottomrule[1.5pt]

    \addlinespace[2pt] 
    \multirow{2}{*}{\textbf{Method}}& \multicolumn{2}{c}{\textbf{CIFAR10}}  & \multicolumn{2}{c}{\textbf{CIFAR100}} & \multicolumn{2}{c}{\textbf{PR}}\\
    
    \cmidrule(lr){2-3} \cmidrule(lr){4-5} \cmidrule(lr){6-7} 
    & \textbf{Res18.} & \textbf{Res34.} & \textbf{Res18.} & \textbf{Res34.}& \textbf{Res18.} & \textbf{Res34.}\\
    \toprule[0.75pt]

    AllSmall & 76.8\% & 67.0\% & 37.5\% & 27.4\% & 100\% & 100\% \\

    ExclusiveFL & 77.9\% & NA & 37.1\% & NA & 18\% & 0\% \\

    DepthFL&79.3\%&80.1\%&36.5\%&47.0\% & 43\% & 36\% \\
    
    HeteroFL&82.8\%&9.9\%&47.0\%&1.1\%  & 100\% & 100\% \\
    
    FedRolex&84.7\%&81.4\%&51.3\%&44.3\%  & 100\% & 100\% \\
    SmartFreeze & 82.8\% & 82.0\% & 54.4\% & 50.7\% & 100\% & 100\%  \\
    ProFL & 85.2\% & 82.6\% & 55.8\% & 52.3\% & 100\% & 100\% \\
    \midrule
    \our&\textbf{87.0\%}&\textbf{84.2\%} & \textbf{57.3\%}&\textbf{54.8\%}  & 100\% & 100\% \\
    \bottomrule[1.5pt]
    \end{tabular}
    \caption{Performance comparison of FL methods on tasks of varying complexity. NA indicates that the algorithm cannot operate under this setup due to memory constraints.}
    \vspace{-4mm}
\label{performance_accvgg}
\end{table}

\noindent\textbf{Robustness to Different Memory Constraints.}  
To evaluate the robustness of \our under varying memory constraints, we conduct experiments with ResNet18 and ResNet34. The increased complexity of ResNet34 imposes more stringent memory requirements on edge devices. Table~\ref{performance_accvgg} presents the experimental results under the IID setting, where we observe that \our consistently delivers superior performance across both models.
It is worth noting that ExclusiveFL becomes infeasible for ResNet34, as no device has enough memory to support model training. Similarly, HeteroFL experiences significant performance drops since no device can update the complete set of channels. In contrast, \our remains effective and achieves up to a 27.4\% accuracy improvement on ResNet34 compared to other baselines. 
These results convincingly demonstrate that \our provides exceptional resilience to memory constraints while maintaining high model performance.

\begin{figure}
    \centering
    \subfigure[FEMNIST.]{
    \includegraphics[width=0.45\linewidth]{ 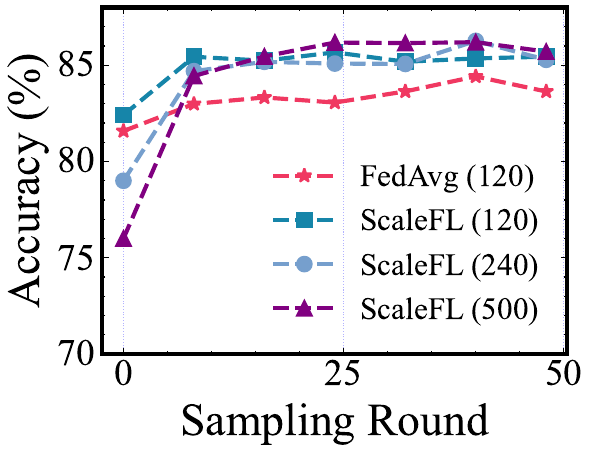}
    \label{femnist1}}
    \subfigure[Vision Transformer.]{
    \includegraphics[width=0.45\linewidth]{ 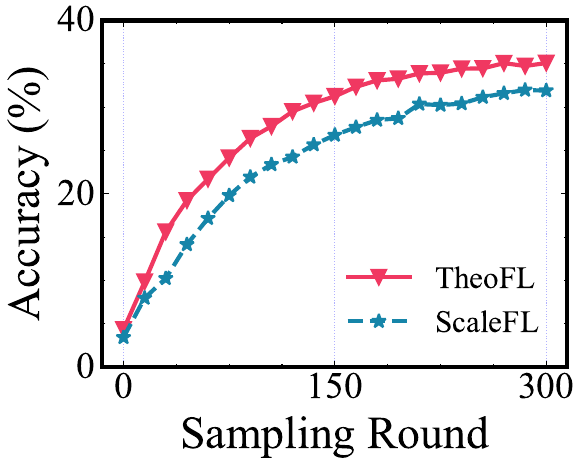}
    \label{femnist2}}
    \vspace{-5pt}
    \caption{Performance comparison on large-scale datasets and Transformer-based models.}
    \label{FEMNIST}
    \vspace{-4mm}
\end{figure}

\noindent\textbf{Scalability with Large-Scale Datasets.}  
To demonstrate the scalability of \our on large-scale datasets, we conduct experiments on the FEMNIST dataset, which contains 805,263 images. 
We evaluate \our across varying numbers of devices (\{120, 240, 500\}), using FedAvg with 120 devices as the baseline.  
The experimental results are shown in Figure~\ref{femnist1}, where the X-axis represents the sampling training rounds and the Y-axis indicates the testing accuracy.  
We observe that \our converges efficiently across different device scales and achieves up to 3\% higher accuracy compared to FedAvg.


\noindent\textbf{Compatibility with Transformer-Based Models.}  
We further evaluate the compatibility of \our with mainstream Transformer-based architectures by conducting experiments on Mini-ImageNet using a ViT configured with 12 Transformer layers as the global model.
Figure~\ref{femnist2} shows the accuracy curves of \our in comparison to the theoretical baseline (\textsc{TheoFL}).  
We observe that \our converges effectively, exhibiting only a modest 2\% accuracy drop compared to \textsc{TheoFL}.
However, it is worth noting that \textsc{TheoFL} is impractical in real-world cases due to memory constraints. These results demonstrate that \our is well-suited for mainstream Transformer-based models. 

\begin{figure}
    \centering
     \subfigure[TX2 (ResNet18).]{
    \includegraphics[width=0.45\linewidth]{ 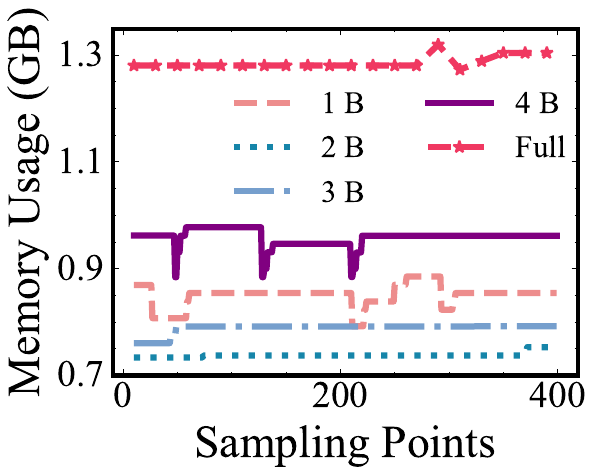}}
    \subfigure[TX2 (ResNet34).]{
    \includegraphics[width=0.45\linewidth]{ 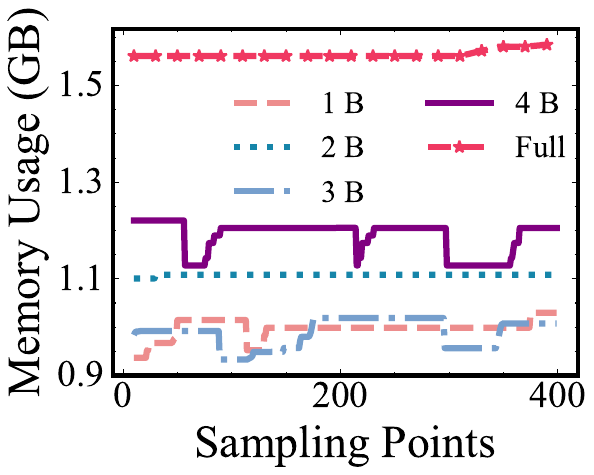}}
    \vspace{-6pt}
    \caption{Memory usage when training different blocks on TX2. \textit{Full} denotes end-to-end training of the entire model.}
    \vspace{-3mm}
    \label{memory_reduction}
\end{figure}

\subsection{Hardware Evaluation}

\noindent\textbf{Memory Efficiency.}  
To assess the memory efficiency of \our on real-world edge devices, we measure memory usage during training ResNet18 and ResNet34 on a Jetson TX2.  
Figure~\ref{memory_reduction} illustrates the memory footprint of training individual blocks using \our, in comparison to end-to-end training of the entire model.  
The X-axis represents sampling points during local training, while the Y-axis shows the  memory usage.  
We observe that \our significantly reduces peak memory usage by training only one block per round.  
Excluding the memory usage of the process itself, \our reduces peak memory usage by up to 50.4\%. 

\definecolor{red}{RGB}{172,21,28}
\definecolor{blue}{RGB}{39,89,167}
\definecolor{red1}{RGB}{203,104,104}
\definecolor{blue1}{RGB}{104,155,203}
\definecolor{color1}{RGB}{235,164,122}
\definecolor{color3}{RGB}{78,172,183}
\definecolor{color2}{HTML}{b5cfd8}
\definecolor{color4}{HTML}{f0bebe}
\definecolor{color5}{HTML}{8aae92}

\begin{figure}[t]
    \centering
    \vspace{1.0mm}
\begin{tikzpicture}
    \scriptsize{

  \begin{axis}[
    at={(-5.5em,-15.5em)},
    anchor=south west,
    ymajorgrids,
    grid style=dashed,
    legend style={at={(-0.15,1)}, anchor=south west},
    legend cell align={left},
    ybar,
    enlarge x limits=0.5,
    xtick align=inside,
    height=.23\textwidth,
    width=.26\textwidth,
    bar width=0.8em,
    xlabel={\scriptsize{(a) TX2 (CIFAR10).}},
    xlabel style={scale=1.2, yshift=0.3em, xshift=0.1em},
    ylabel=\footnotesize{\scriptsize Training Time (S)},
    ylabel style={scale=1.2, yshift=0.1em},
    symbolic x coords={{1}, {2},},
    xtick=data,
    ymin=0,
    ymax=130,
    ytick={0,40,80,120},
    nodes near coords align={vertical},
    xticklabels={ResNet18, ResNet34},
    ylabel style={yshift=-2em},
    yticklabel style={/pgf/number format/fixed,/pgf/number format/fixed zerofill,/pgf/number format/precision=0,rotate=0,scale=1.0},
    legend style={yshift=0.2em,xshift=4.2em,font={\tiny},cells={anchor=west},fill opacity=0.8, scale=1.0, legend columns=5}
    ]
    \addplot[fill=color1, draw=color1, area legend] coordinates {({1},29.90204906463623) ({2},40.30962419509888)};
    \addlegendentry{\scalebox{1.0}{1 B}}
    \addplot[fill=color2, draw=color2, area legend] coordinates {({1},29.013141870498657) ({2},52.1981463432312)};
    \addlegendentry{\scalebox{1.0}{2 B}}
    \addplot[fill=color3, draw=color3, area legend] coordinates {({1},28.97889018058777) ({2},66.88013482093811)};
    \addlegendentry{\scalebox{1.0}{3 B}}
    \addplot[fill=color4, draw=color4, area legend] coordinates {({1},36.22412919998169) ({2},71.4707574844360)};
    \addlegendentry{\scalebox{1.0}{4 B}}
    \addplot[fill=color5, draw=color5, area legend] coordinates {({1},66.61795830726624) ({2},125.03500699996948)};
    \addlegendentry{\scalebox{1.0}{Full}}
  \end{axis}

  \begin{axis}[
    at={(10.5em,-15.5em)},
    anchor=south west,
    ymajorgrids,
    grid style=dashed,
    legend style={at={(-0.15,1)}, anchor=south west},
    legend cell align={left},
    ybar,
    enlarge x limits=0.5,
    xtick align=inside,
    height=.23\textwidth,
    width=.26\textwidth,
    bar width=0.8em,
    xlabel={\scriptsize{(b) TX2 (CINIC10).}},
    xlabel style={scale=1.2, yshift=0.3em, xshift=0.1em},
    ylabel=\footnotesize{\scriptsize Training Time (S)},
    ylabel style={scale=1.2, yshift=0.1em},
    symbolic x coords={{1}, {2},},
    xtick=data,
    ymin=0,
    ymax=260,
    ytick={0,80,160,240},
    nodes near coords align={vertical},
    xticklabels={ResNet18, ResNet34},
    ylabel style={yshift=-2em},
    yticklabel style={/pgf/number format/fixed,/pgf/number format/fixed zerofill,/pgf/number format/precision=0,rotate=0,scale=1.0},
    legend style={yshift=0.2em,xshift=4.2em,font={\tiny},cells={anchor=west},fill opacity=0.8, scale=1.0, legend columns=5}
    ]
    \addplot[fill=color1, draw=color1, area legend] coordinates {({1},53.96686267852783) ({2},70.54075)};
    \addplot[fill=color2, draw=color2, area legend] coordinates {({1},52.222087144851685) ({2},88.7368)};
    \addplot[fill=color3, draw=color3, area legend] coordinates {({1},51.576401710510254) ({2},113.69)};
    \addplot[fill=color4, draw=color4, area legend] coordinates {({1},64.61289811134338) ({2},122.069)};
    \addplot[fill=color5, draw=color5, area legend] coordinates {({1},120.08382487297058) ({2},222.55)};
  \end{axis}
}   
\end{tikzpicture}
\vspace{-3mm}
\caption{Per-round training time for individual blocks.}
\label{fig:efficiency}
\vspace{-4mm}
\end{figure}

\noindent\textbf{Training Efficiency.} 
\our is also an efficient training framework, as the training time per round is significantly reduced compared to end-to-end training. 
To demonstrate this, we experiment with ResNet18 on the CIFAR10 and CINIC10 datasets, tracking the training time for each block on a Jetson TX2. 
As shown in Figure~\ref{fig:efficiency}, training individual blocks reduces per-round training time by 1.84$\times$ to 2.31$\times$ relative to end-to-end training of the entire model.  
When applied to ViT, \our achieves up to a 1.9$\times$ speedup over \textsc{TheoFL}, further highlighting its training efficiency.

\definecolor{red}{RGB}{172,21,28}
\definecolor{blue}{RGB}{39,89,167}
\definecolor{red1}{RGB}{203,104,104}
\definecolor{blue1}{RGB}{104,155,203}
\definecolor{color1}{HTML}{669b7c}
\definecolor{color2}{HTML}{7db9b3}
\definecolor{color3}{HTML}{7aa5d2}
\definecolor{color4}{HTML}{f6dec4}

\begin{figure}[t]
    \centering
    \vspace{1.0mm}
\begin{tikzpicture}
    \scriptsize{

  \begin{axis}[
    at={(-5.5em,-15.5em)},
    anchor=south west,
    ymajorgrids,
    grid style=dashed,
    legend style={at={(-0.2,1)}, anchor=south west},
    legend cell align={left},
    ybar,
    enlarge x limits=0.5,
    xtick align=inside,
    height=.23\textwidth,
    width=.26\textwidth,
    bar width=1em,
    xlabel={\scriptsize{(a) CIFAR10.}},
    xlabel style={scale=1.2, yshift=0.3em, xshift=0.1em},
    ylabel=\footnotesize{\scriptsize Accuracy (\%)},
    ylabel style={scale=1.2, yshift=0.1em},
    symbolic x coords={{1}, {2},},
    xtick=data,
    ymin=70,
    ymax=95,
    ytick={70,80,90},
    nodes near coords align={vertical},
    xticklabels={IID, non-IID},
    ylabel style={yshift=-2em},
    yticklabel style={/pgf/number format/fixed,/pgf/number format/fixed zerofill,/pgf/number format/precision=0,rotate=0,scale=1.0},
    legend style={yshift=0.2em,xshift=4.2em,font={\tiny},cells={anchor=west},fill opacity=0.8, scale=1.0, legend columns=5}
    ]
    \addplot[fill=color1, draw=color1, area legend] coordinates {({1},87.0) ({2},80.4)};
    \addlegendentry{\scalebox{1.0}{\our}}
    \addplot[fill=color2, draw=color2, area legend] coordinates {({1},86.5) ({2},78.6)};
    \addlegendentry{\scalebox{1.0}{w/o CM}}
    \addplot[fill=color3, draw=color3, area legend] coordinates {({1},79.6) ({2},77.0)};
    \addlegendentry{\scalebox{1.0}{w/o TH}}
    \addplot[fill=color4, draw=color4, area legend] coordinates {({1},77.9) ({2},76.8)};
    \addlegendentry{\scalebox{1.0}{FedAvg}}
  \end{axis}

  \begin{axis}[
    at={(10.5em,-15.5em)},
    anchor=south west,
    ymajorgrids,
    grid style=dashed,
    legend style={at={(-0.15,1)}, anchor=south west},
    legend cell align={left},
    ybar,
    enlarge x limits=0.5,
    xtick align=inside,
    height=.23\textwidth,
    width=.26\textwidth,
    bar width=1em,
    xlabel={\scriptsize{(b) CIFAR100.}},
    xlabel style={scale=1.2, yshift=0.3em, xshift=0.1em},
    ylabel=\footnotesize{\scriptsize Accuracy (\%)},
    ylabel style={scale=1.2, yshift=0.1em},
    symbolic x coords={{1}, {2},},
    xtick=data,
    ymin=30,
    ymax=65,
    ytick={30,45,60},
    nodes near coords align={vertical},
    xticklabels={IID, non-IID},
    ylabel style={yshift=-2em},
    yticklabel style={/pgf/number format/fixed,/pgf/number format/fixed zerofill,/pgf/number format/precision=0,rotate=0,scale=1.0},
    legend style={yshift=0.2em,xshift=4.2em,font={\tiny},cells={anchor=west},fill opacity=0.8, scale=1.0, legend columns=5}
    ]
    \addplot[fill=color1, draw=color1, area legend] coordinates {({1},57.3) ({2},51.2)};
    \addplot[fill=color2, draw=color2, area legend] coordinates {({1},56.1) ({2},50.5)};
    \addplot[fill=color3, draw=color3, area legend] coordinates {({1},48.3) ({2},42.3)};
    \addplot[fill=color4, draw=color4, area legend] coordinates {({1},37.1) ({2},35.2)};
  \end{axis}
}   
\end{tikzpicture}
\vspace{-3mm}
\caption{Ablation Study of \our. {w/o~CM} indicates the Curriculum Mentor is removed, and {w/o~TH} signifies the Training Harmonizer is removed.}
\label{fig:ablation}
\vspace{-5mm}
\end{figure}

\subsection{Ablation Study}

Finally, we conduct a breakdown analysis of the benefits brought by each component, i.e., the Curriculum Mentor and the Training Harmonizer. We perform experiments with ResNet18 on the CIFAR10 and CIFAR100 datasets. Figure~\ref{fig:ablation} shows that both the Curriculum Mentor and the Training Harmonizer have non-trivial contributions to performance improvement. For example, on the CIFAR100 dataset under the IID condition, the accuracy declines by 1.2\% without the Curriculum Mentor and significantly drops by 9.0\% without the Training Harmonizer. However, there is still an accuracy improvement of 19\% and 11.2\% over FedAvg, respectively. These results confirm the effectiveness of these components and the sequential block-wise training paradigm. 
\section{Related Work}

\noindent\textbf{Resource-Constrained FL.} 
FedMD~\cite{li2019fedmd}, a model-heterogeneous training method, customizes local models based on each device’s memory capacity and enables clients to share logits on a public dataset to facilitate knowledge transfer. However, obtaining such a high-quality public dataset is challenging in FL due to privacy concerns~\cite{zhang2022fedzkt}. 
To address this, HeteroFL~\cite{diao2020heterofl} and FedRolex~\cite{alam2022fedrolex} customize sub-models of varying complexity for each device by scaling the number of channels in convolutional layers.
However, scaling the model width compromises the model architecture and leads to degraded performance. Therefore, DepthFL~\cite{kim2022depthfl} and InclusiveFL~\cite{liu2022no} scale the model along the depth dimension to produce sub-models of different complexity.
However, these methods typically assume the existence of at least one device that can train the entire model, which inherently bounds the global model complexity. In contrast, \our trains the global model in a block-wise manner, effectively overcoming the limitations of the aforementioned methods.

\noindent\textbf{Block-wise Training.}  
ProgFed~\cite{wang2022progfed} partitions the global model into blocks and progressively introduces new blocks for training. However, it still requires end-to-end training of the entire model and therefore fails to effectively address the memory constraints. Unlike ProgFed, SmartFreeze~\cite{wu2024heterogeneity} trains each block sequentially. After a block converges, it is frozen and the next block is introduced. This intelligent freezing strategy significantly reduces peak memory usage, enabling devices with limited memory to participate. ProFL~\cite{wu2025breaking} further decomposes the training process into shrinking and growing stages to better support the sequential block-wise training paradigm. However, these methods overlook the challenges of information loss and inter-block information isolation that arise during block-wise training. \our is specifically designed to address these issues effectively.

\vspace{-1mm}
\section{Conclusion}
This paper introduces \our, a scalable and inclusive framework that addresses the memory constraints of participating devices via sequential block-wise training. \our integrates two core components to guide the training process of each block.
Specifically, the Curriculum Mentor crafts curriculum-aware training objectives for each block to mitigate information loss. The Training Harmonizer implements a parameter co-adaptation training scheme to break information isolation across blocks. 
Extensive experiments are conducted to demonstrate the effectiveness of \our.

\bibliography{ref}

\section{Formulation of HSIC}\label{appendix_HSIC}

The Hilbert-Schmidt Independence Criterion (HSIC) quantifies the statistical dependence between two random variables \( X \) and \( Z \) using kernel embeddings. It is defined as:
\begin{equation}
\small
\begin{aligned}
    \mathrm{HSIC}(X, Z) &= \mathbb{E}_{X,Z,X',Z'} \left[ k_X(X, X') \cdot k_Z(Z, Z') \right] \\
    &\quad + \mathbb{E}_{X,X'} \left[ k_X(X, X') \right] \cdot \mathbb{E}_{Z,Z'} \left[ k_Z(Z, Z') \right] \\
    &\quad - 2 \cdot \mathbb{E}_{X,Z} \left[ \mathbb{E}_{X'} \left[ k_X(X, X') \right] \cdot \mathbb{E}_{Z'} \left[ k_Z(Z, Z') \right] \right],
\end{aligned}
\end{equation}
where \( k_X \) and \( k_Z \) are positive-definite kernel functions applied to independent copies \( (X, X') \) and \( (Z, Z') \), respectively. Expectations are taken over the joint distributions of the corresponding variables.

\section{Convergence Analysis of \our}\label{appendix_convergence}

In this section, we present a convergence analysis for \our. Our proof follows the standard methodology in federated learning theory~\cite{li2019convergence, li2022one, wu2025breaking, wang2022progfed}, with extensions to accommodate the sequential block-wise training paradigm, curriculum-aware training objectives, and the parameter co-adaptation scheme. We demonstrate that, under common assumptions (e.g., smoothness and bounded gradients), the sequence of global parameters \(\{\Theta_{g,t}^r\}\) converges to a stationary point.

\subsection{Preliminaries and Assumptions}
\begin{enumerate}
    \item \textbf{Smoothness:} Each local objective function is \(L\)-smooth. Formally, for each client \(n\), we assume the loss \(\mathcal{L}^{r}_{n,t}\) in Equation~\eqref{IB_loss2} satisfies:
    \begin{equation}
    \label{equ:lsmooth}
        \bigl\|\nabla \mathcal{L}_{n,t}^{r}(\theta) - \nabla \mathcal{L}_{n,t}^{r}(\theta')\bigr\| \leq L \bigl\|\theta - \theta'\bigr\|.
    \end{equation}
    \item \textbf{Bounded Gradients:} The gradient norms in each training round are uniformly bounded, i.e., \(\bigl\|\nabla \mathcal{L}_{n,t}^{r}(\theta)\bigr\| \leq G\) for some \(G > 0\).
    \item \textbf{Data Heterogeneity:} While federated data may be non-IID, we account for this via the regularization term in Equation~\eqref{IB_loss2}, which bounds client drift.
    \item \textbf{Sequential Block-Wise Training:} The global model \(\Theta\) is divided into blocks \([\theta_{1}, \theta_{2}, \ldots, \theta_{T}]\). At stage \(t\), only the parameters \(\theta_{t}\) (along with a small set of connecting layers for backward gradient flow) are updated locally and then aggregated as follows:
    \begin{equation}
    \label{eq_appendix_aggre}
        [L_{g,t-1}^{r}, \theta_{g,t}^{r}, \theta_{g,Op}^{r}] = \sum_{n \in S} \frac{|D_{n}|}{|D|}([L_{n,t-1}^{r}, \theta_{n,t}^{r}, \theta_{n,Op}^{r}]).
    \end{equation}
\end{enumerate}

\subsection{Main Theorem}
\begin{theorem}
\label{thm:convergence}

Under the smoothness and bounded-gradient assumptions above, let \(\{\Theta_{g,t}^r\}\) be the sequence of global models produced by \our, where each block \(\theta_t\) is updated using the curriculum-aware objective in Equation~\eqref{IB_loss2}, and aggregated via Equation~\eqref{eq_appendix_aggre} with a sufficiently small learning rate. Then, there exists a constant \(\Psi > 0\) such that:
\begin{equation}
    \frac{1}{R} \sum_{r=1}^{R} \E\bigl[\bigl\|\nabla \mathcal{L}_{t}^{r}(\Theta_{g,t}^r)\bigr\|^2\bigr] \leq \frac{\Psi}{\sqrt{R}},
\end{equation}
where \(R\) is the total number of rounds. Consequently, the average gradient norm converges to \(0\), implying that \(\{\Theta_{g,t}^r\}\) converges to a stationary point.

\end{theorem}

\subsection{Proof}
\begin{proof}
Our proof proceeds by induction over blocks and rounds.

\paragraph{Step 1: Block-Wise Gradient Analysis.}
From the local training objective in Equation~\eqref{IB_loss2}, block \(\theta_t\) at each training round \(r\) involves minimizing:
\begin{equation}
    \small
    \mathcal{L}_{t}^{r} = \mathcal{L}_{CE} - \lambda_{t} \cdot \text{nHSIC}(X; Z_{t}) - \gamma_{t} \cdot \text{nHSIC}(Y; Z_{t})
    + \frac{\mu}{2}\bigl\|\theta_{t}^{r} - \theta_{t}^{r-1}\bigr\|_2^2.
\end{equation}
Due to \(L\)-smoothness as shown in Equation~\eqref{equ:lsmooth}, we can bound the decrease of \(\mathcal{L}_{t}^{r}\) by the gradient norms with respect to \(\theta_{t}^r\). Given that only \(\theta_{t}^r\) (along with a small set of parameters from the connecting layers) is updated in each round, we have:
\begin{equation}
    \E\Bigl[\mathcal{L}_{t}^{r}(\theta_{t}^r) - \mathcal{L}_{t}^{r}(\theta_{t}^{r+1})\Bigr]
    \;\;\ge\;\; \frac{\eta}{2}\E\bigl[\bigl\|\nabla \mathcal{L}_{t}^{r}(\theta_{t}^r)\bigr\|^2\bigr] - \eta^2 C,
\end{equation}
where \(\eta\) is the local learning rate and \(C\) is a constant proportional to the square of the bounded gradients.

\paragraph{Step 2: One-Round Analysis.}
At the end of each round, parameter updates from clients in set \(S\) are aggregated via Equation~\eqref{eq_appendix_aggre}.
Following the standard FedAvg analysis~\cite{li2019convergence}, we can derive that:
\begin{equation}
    \E\bigl[\|\theta_{g,t}^{r+1} - \theta_{g,t}^{r}\|^2\bigr]
    \;\;\le\;\; \eta^2 L^2 \E\bigl[\|\nabla \mathcal{L}_{t}^{r}\|^2\bigr] + \text{(client drift terms)},
\end{equation}
and thus the global model shift can be made arbitrarily small with a suitable choice of \(\eta\) and our regularization.

\paragraph{Step 3: Impact of Sequential Training.}
The sequential training paradigm of \our ensures that each block \(\theta_{t}\) benefits from consecutive re-training steps, as well as from the concurrent updates of preceding blocks by the Training Harmonizer (i.e., partial unfreezing). This co-adaptation mechanism helps to reduce error propagation across blocks. Additionally, the inclusion of
\(\lambda_{t}\,\text{nHSIC}(X; Z_{t})\) and \(\gamma_{t}\,\text{nHSIC}(Y; Z_{t})\) in Equation~\eqref{IB_loss_old} 
guarantees that each block neither discards too much input information nor fails to attend to task-relevant features.

\paragraph{Step 4: Multi-Round Convergence.}
Summing the per-round descent bounds over \(r=1\) to \(R\) and applying Jensen’s inequality~\cite{pap2010generalization} to the expected gradient term yields:
\begin{equation}
    \sum_{r=1}^{R}\E\bigl[\bigl\|\nabla \mathcal{L}_{t}^{r}(\Theta_{g,t}^r)\bigr\|^2\bigr]
    \;\;\le\;\; \Tilde{C}\,\sqrt{R} \;+\; O(\eta^2),
\end{equation}
for some constant \(\Tilde{C} > 0\). Dividing both sides by \(R\) and letting \(R \to \infty\) confirms that:
\begin{equation}
    \frac{1}{R}\sum_{r=1}^{R}\E\bigl[\bigl\|\nabla \mathcal{L}_{t}^{r}(\Theta_{g,t}^r)\bigr\|^2\bigr] \to 0,
\end{equation}
which implies convergence to a stationary point. Therefore, as each block and its partial outputs converge, the entire model \(\Theta\) also converges under \our’s training paradigm.
  \renewcommand{\qed}{}  
\end{proof}

\paragraph{Conclusion.}
The Theorem \ref{thm:convergence} shows that \our converges under standard assumptions. The key elements driving this result are: 
1) the \emph{curriculum-aware training objective}, which prevents severe information loss; 
2) the \emph{parameter co-adaptation training scheme}, which breaks information isolation across blocks; and 
3) bounded and smooth loss landscapes. 
Together, these ensure that the sequential block-wise training paradigm maintains gradient alignment and preserves salient features, thus achieving stable convergence.





\setlength{\leftmargini}{20pt}
\makeatletter\def\@listi{\leftmargin\leftmargini \topsep .5em \parsep .5em \itemsep .5em}
\def\@listii{\leftmargin\leftmarginii \labelwidth\leftmarginii \advance\labelwidth-\labelsep \topsep .4em \parsep .4em \itemsep .4em}
\def\@listiii{\leftmargin\leftmarginiii \labelwidth\leftmarginiii \advance\labelwidth-\labelsep \topsep .4em \parsep .4em \itemsep .4em}\makeatother

\setcounter{secnumdepth}{0}
\renewcommand\thesubsection{\arabic{subsection}}
\renewcommand\labelenumi{\thesubsection.\arabic{enumi}}

\newcounter{checksubsection}
\newcounter{checkitem}[checksubsection]

\newcommand{\checksubsection}[1]{%
  \refstepcounter{checksubsection}%
  \paragraph{\arabic{checksubsection}. #1}%
  \setcounter{checkitem}{0}%
}

\newcommand{\checkitem}{%
  \refstepcounter{checkitem}%
  \item[\arabic{checksubsection}.\arabic{checkitem}.]%
}
\newcommand{\question}[2]{\normalcolor\checkitem #1 #2 \color{blue}}
\newcommand{\ifyespoints}[1]{\makebox[0pt][l]{\hspace{-15pt}\normalcolor #1}}

\end{document}